\def\year{2022}\relax
\documentclass[letterpaper]{article} 
\usepackage{aaai21}  
\usepackage{times}  
\usepackage{helvet} 
\usepackage{courier}  
\usepackage[hyphens]{url}  
\usepackage{graphicx} 
\usepackage{csquotes}
\usepackage{amssymb}
\usepackage{rotating}
\usepackage{amsmath}
\usepackage{natbib}  
\usepackage{caption} 
\usepackage{array}
\usepackage{hhline}
\usepackage{flushend}
\usepackage{setspace} 
\usepackage{makecell}
\usepackage[flushleft]{threeparttable}
\usepackage{booktabs,caption}
\usepackage[toc,page]{appendix}
\usepackage{subcaption}
\usepackage{rotating}
\usepackage{multirow}
\usepackage[T1]{fontenc}

\title{American Twitter Users Revealed Social Determinants-related Oral Health Disparities amid the COVID-19 Pandemic}
\author {
    Yangxin Fan, \textsuperscript{\rm 1}
    Hanjia Lyu, \textsuperscript{\rm 2}
    Jin Xiao, \textsuperscript{\rm 3}
    Jiebo Luo \textsuperscript{\rm 2}\\
}
\affiliations{
    \textsuperscript{\rm 1} Goergen Institute for Data Science, University of Rochester\\
    \textsuperscript{\rm 2} Department of Computer Science, University of Rochester\\
    \textsuperscript{\rm 3} Eastman Institute for Oral Health, University of Rochester Medical Center\\
    \{yfan24, hlyu5\}@ur.rochester.edu, jin\_xiao@urmc.rochester.edu, jluo@cs.rochester.edu\\}
\begin{document}


\maketitle

\begin{abstract}
\textbf{Objectives:} To assess self-reported population oral health conditions amid COVID-19 pandemic using user reports on Twitter.

\textbf{Method and Material:} We collected oral health-related tweets during the COVID-19 pandemic from 9,104 Twitter users across 26 states (with sufficient samples) in the United States between November 12, 2020 and June 14, 2021. We inferred user demographics by leveraging the visual information from the user profile images. Other characteristics including income, population density, poverty rate, health insurance coverage rate, community water fluoridation rate, and relative change in the number of daily confirmed COVID-19 cases were acquired or inferred based on retrieved information from user profiles. We performed logistic regression to examine whether discussions vary across user characteristics.

\textbf{Results:} Overall, 26.70\% of the Twitter users discuss wisdom tooth pain/jaw hurt, 23.86\% tweet about dental service/cavity, 18.97\% discuss chipped tooth/tooth break, 16.23\% talk about dental pain, and the rest are about tooth decay/gum bleeding. Women and younger adults (19-29) are more likely to talk about oral health problems. Health insurance coverage rate is the most significant predictor in logistic regression for topic prediction.

\textbf{Conclusion:} Tweets inform social disparities in oral health during the pandemic. For instance, people from counties at a higher risk of COVID-19 talk more about tooth decay/gum bleeding and chipped tooth/tooth break. Older adults, who are vulnerable to COVID-19, are more likely to discuss dental pain. Topics of interest vary across user characteristics. Through the lens of social media, our findings may provide insights for oral health practitioners and policy makers.

\end{abstract}

\section{Introduction}
After the World Health Organization declared the global spread of COVID-19 a pandemic on March 11, 2020,\footnote{\url{https://pubmed.ncbi.nlm.nih.gov/32191675/}} lockdowns were enforced nationwide in the US to reduce the spread of the virus. At the early outbreak of the COVID-19 pandemic, the American Dental Association (ADA) recommended that dental practices postpone elective procedure and provide emergency-only dental services.\footnote{\url{https://www.ada.org/en/publications/ada-news/2020-archive/march/ada-recommending-dentists-postpone-elective-procedures}} As a result, patients’ access to dental services have greatly decreased. During the week of March 23, 2020, an ADA Health Policy Survey indicated that 19\% of dental offices were completely closed and 76\% were partly closed but seeing emergency patients only~\cite{american2020hpi}. More importantly, loss of dental insurance by many people has also increased the risk of oral diseases. According to a survey commissioned by the CareQuest Institute for Oral Health,\footnote{\url{https://www.carequest.org/system/files/CareQuest-Institute-Coming-Surge-Oral-Health-Treatment-Needs-Report-1.pdf}} an estimated six million American adults have lost their dental insurance and 28 million American adults delayed getting dental care. Although most dental clinics reopened in June 2020, dental services have not rebounded to the full capacity due to office infection control regulation, lack of PPEs, and reduced patient-initiated dental visits.\footnote{\url{https://www.ada.org/en/science-research/health-policy-institute/covid-19-dentists-economic-impact/survey-results}} Additionally, studies have found that there is an association between oral health and severity of COVID-19 complications that makes preventing bad oral health even more challenging~\cite{sampson2020could,botros2020there}. An indirect connection has also been suggested that due to the work from home (WFH) policy, people increase consumption of products that are likely to be detrimental to dental health (e.g., alcohol, sweets) but also increase the use of products which benefit oral health.\footnote{\url{https://bridgedental.com/2020/04/26/the-impact-of-working-from-home-during-the-coronavirus-lockdown-on-dental-health/}}

Given the emergency of dental care caused by lack of access to dental services and loss of dental insurance, we attempted to identify the vulnerable groups of people by analyzing factors like age, gender, population density, income, poverty rate, health insurance coverage, community water fluoridation, as well as relative change in the number of daily confirmed COVID-19 cases. In addition, we intended to explore what kinds of oral diseases or issues to which they were more likely to be exposed during the COVID-19 pandemic. To our best knowledge, this is the first large-scale social media-based study to analyze and understand oral health in America amid the COVID-19 pandemic. We hope our study through the lens of social media, especially the findings of social disparities, can provide insights for oral health practitioners and policy makers.

Oral health services in the US face an unprecedented challenge during the COVID-19 outbreak. On one hand, the COVID-19 pandemic increased the risk for oral diseases of whom are vulnerable to COVID-19, including those in rural areas, low socioeconomic groups, older adults, disadvantaged and underprivileged children, and the uninsured~\cite{kalash2020covid,marchini2020covid,zachary2020oral}. On the other hand, complex effect from intensified COVID-19 therapies and multi-drug treatment could possibly further exacerbate some oral conditions~\cite{dziedzic2021impact}. COVID-19 also has direct effects on oral health through its official symptom ageusia~\cite{daly2020impact}. Despite of the fact that COVID-19 testing positivity rates were low among practicing US dentists~\cite{estrich2020estimating}, the fear of contacting the virus may still lead to resistance to dental treatment, which in turn will increase of the level of dental anxiety~\cite{gonzalez2020perceived}.

Although the COVID-19 pandemic has greatly impacted people’s oral health over the last year, as an innovative way of disease diagnosis, telemedicine has gained public attention since it has the potential to provide the patients with the clinical care they need while remaining the distance~\cite{hollander2020virtually}. An example of telemedicine for oral health is to utilize the instant text and image messaging functions from social media platforms to diagnose and counsel for oral diseases in the COVID-19 era~\cite{machado2020social}. Many large-scale social media-based studies have investigated different public health topics amid the COVID-19 pandemic, such as acquiring insights about the US mental health during the COVID-19 pandemic from Twitter data~\cite{valdez2020social}, studying the nature and diffusion of COVID-19 related oral health information using tweets from Chinese social media Weibo~\cite{tao2020nature}, monitoring depression trends on Twitter during the COVID-19 pandemic~\cite{zhou2021detecting,zhang2021monitoring}, tracking mental health~\cite{valdez2020social,guntuku2020tracking}, and investigating public opinions on COVID-19 vaccines~\cite{lyu2022social,bonnevie2021quantifying,wu2021characterizing}.

Twitter has been a popular social media platform for people in the US to express their views and share their lives with each other. As of July 2021, there are about 73 million Twitter users in the US.\footnote{\url{https://www.statista.com/statistics/242606/number-of-active-twitter-users-in-selected-countries/}} In this study, we intended to understand online discussions on oral health during the COVID-19 pandemic. We conducted a large-scale social media-based study of 9,104 Twitter users across 26 states (with sufficient samples) in the US for the period between November 12, 2020 to June 14, 2021. We collected our data using Tweepy\footnote{\url{https://www.tweepy.org/}} and acquired or inferred user characteristics based on the publicly available information of Twitter users. Particularly, our study aims to answer the following research questions:
\begin{itemize}
    \item \textbf{RQ1:} What are the major topics/oral diseases discussed in oral health-related tweets among American Twitter users?
    \item \textbf{RQ2:} How does discussion of each type of topic/oral disease vary across user characteristics including age, gender, population density, income and poverty rate?
    \item \textbf{RQ3:} How does health insurance coverage rate, relative change in the number of daily confirmed COVID-19 cases, and community water fluoridation rate influence users’ probability of tweeting about different topics/oral diseases?
\end{itemize}

To summarize, we made the following three major contributions:
\begin{itemize}
    \item By applying Latent Dirichlet Allocation (LDA) topic modeling~\cite{blei2003latent}, our study discovered five major topics/oral diseases, including dental pain, dental service/cavity, tooth decay/gum bleeding, wisdom tooth pain/jaw hurt, and chipped tooth/tooth break.
    \item By conducting multiple logistic regression analyses, we found that discussions of topics/diseases vary across user demographics.
    \item Our analyses showed social disparities in oral health that people from the counties with higher health insurance coverage rate tend to tweet less about oral diseases in general and people from counties at a higher risk of COVID-19 tend to tweet less about dental service/cavity but more about oral diseases like tooth decay/gum bleeding and chipped tooth/tooth break. Older people mention dental pain more frequently.
\end{itemize}

\section{Method and Material}
In this section, we introduce the data collection and the methods we used in our analyses. To address RQ1, we discuss how we extracted topics using Topic Modeling. To investigate RQ2 and RQ3, we describe how we inferred the user characteristics and conducted the logistic regressions.

\subsection{Data Collection and Preprocessing}
We collected oral health-related tweets through Tweepy using a list of keywords including “tooth decay”, “cavity”, “black hole”, “food stuck on teeth”, “gums bleeding”, “gums red”, “gums inflammation”, “face swelling”, “cheek swelling”, “drain in my mouth”, “tongue swelling”, “cannot swallow”, “tooth chipped”, “tooth break”, “pain”, “throbbing”, “radiate to the ear”, “jaw hurts”, “can’t open the mouth”, and “wisdom tooth hurts”. However, simply using keyword search may collect many false positive tweets. In particular, the tweets with only “pain”, “black hole”, “cavity”, or “throbbing” may not be related to dental health. Therefore, we removed this kind of tweets by adding one constraint. If the tweet only contains one of the keywords “pain”, “black hole”, “cavity”, or “throbbing” but does not contain any other keywords from the aforementioned keyword list, this tweet is excluded from our dataset. To validate this method, we randomly sampled 1,000 tweets from the tweet pool after excluding irrelevant tweets (we read them to examine if they are indeed related to our study). Using our method, 1.7\% are labeled as relevant and 94.1\% of them are indeed related to dental health discussions. Of the 98.3\% tweets that are labeled as irrelevant, none of them are actually related to dental health discussions. To verify whether our method can filter out oral health advertisements, we randomly sampled 500 tweets. Among these sampled tweets, only 1.8\% are oral health advertisements and are mostly from dentists. Our validations indicate the high performance of our excluding criteria for both irrelevant tweets and oral advertisements. In addition, since our study focused on understanding the online discussions of US Twitter users, we excluded the tweets that were not posted by the users whose profile indicated a US location. After removing the irrelevant tweets, the dataset is composed of 21,677 tweets for the period between November 12, 2020 to June 14, 2021 tweeted by 15,133 unique users.

\subsection{Feature Inference}
\subsubsection{Age and Gender}
Following the methods used in \citet{lyu2020sense}, we used the Face++ API\footnote{\url{https://www.faceplusplus.com/}} to infer the age and gender information of Twitter users based on the visual information from their profile images.  Face++ detected faces in images and leveraged deep-learning based recognition algorithms to analyze face attributes including age and gender. Users may upload pictures for their profile images. There may be multiple faces in one profile image. To achieve the most robust inference of the demographic information of the Twitter users, we only included users with one intelligible face. In addition, the invalid image URLs were removed. Face++ is one of the most robust image-based inference method with respect to age and gender inference~\cite{jung2018assessing}. \citet{jung2018assessing} evaluated the performance of Face++ on gender and age inference by comparing the machine-inferred labels with the human-annotated labels. They found that Face++ achieved a good performance and matched the human annotations well. There might potentially be users who did not identify themselves as either male or female. Following the designs of most previous studies in the field of dental health~\cite{stouthard1993assessment,cohen2000impact}, we only focused on the male and female users and framed the gender as a binary variable.

Age was binned into five groups: $\leq$ 18, 19-29, 30-49, 50-64, and $\geq$ 65. Users who are younger than 18 years old, between 19 to 29, between 30 to 49, and between 50 to 64 years old account for 1.6\%, 48.6\%, 37.3\%, and 9.3\%, respectively. The rest are over or equal to 65 years old. According to a report from the Pew Research Center~\cite{wojcik2019sizing}, among the US adult Twitter users, 29\% are between 18-29 years old, 44\% are between 30-49 years old, 19\% are between 50-64 years old, and 8\% are over or equal to 65 years old. Compared to the age distribution of general Twitter users, there are proportionally more adults between 19 to 29 years old in our dataset. This is \textit{consistent} with the finding of a household survey\footnote{\url{https://www.ada.org/~/media/ADA/Science\%20and\%20Research/HPI/OralHealthWell-Being-StateFacts/US-Oral-Health-Well-Being.pdf}} that younger adults are most likely to report problems regarding the condition of their mouth and teeth. With respect to gender, as of January 2021, the gender distribution of Twitter users in the US is biased towards men who account for 61.6\% of total users.\footnote{\url{https://www.statista.com/statistics/678794/united-states-twitter-gender-distribution/}} However, in our dataset, 57.4\% users are women. Women tend to tweet about dental health more actively which echoes the previous study that women are more dentally anxious~\cite{doerr1998factors}.

\subsubsection{Population Density}
We applied a Python package - USZipcode search engine to extract population density of each Twitter user’s location based on their profile information that were reported by themselves. The population density was categorized into three levels: urban (greater than 3,000 people per square mile), suburban (1,000-3,000 people per square mile), and rural (less than 1,000 people per square mile). In our study population, 72.0\% are urban, 12.1\% are suburban, and 15.9\% are rural which is similar to the share reported in a previous report of the Pew Research Center\footnote{\url{https://www.pewresearch.org/fact-tank/2019/04/10/share-of-u-s-adults-using-social-media-including-facebook-is-mostly-unchanged-since-2018/}} that most Twitter users live in urban area. 

\subsubsection{Income and Poverty Rate}
Studies have shown that income or poverty rate is strongly associated with oral cancer, dental caries prevalence, caries experience, and traumatic dental injuries~\cite{singh2019relationship}. Therefore, we propose to include the socioeconomic status of the Twitter users into our study. Specifically, we use the Census API\footnote{\url{https://www.census.gov/data/developers.html}} to retrieve median per capita income and poverty rate at the county level from the 2019 American Community Survey (ACS). 

\subsubsection{Health Insurance Coverage Rate}
To our best knowledge, there is no publicly disclosed detailed information about dental insurance coverage rate at the county or city level. However, \citet{perez2006factors} found that having medical insurance is positively correlated with better dental care coverage. Compared with the people with private health insurance, those who are not insured are more likely to be unable to get dental care~\cite{shi2010access}. The positive association between the health insurance coverage and the accessibility to dental services might be because (1) having health insurance may affect ability to pay for dental services~\cite{slack2007demographic}, and (2) health insurance is consistently related to the use of preventive medical and dental care~\cite{stella2001factors}. Thus, we chose to use the health insurance coverage rate to approximately measure each user’s accessibility to dental services. We used the Census API to retrieve health insurance coverage at the county level from the 2019 American Community Survey (ACS). 

\subsubsection{Fluoridation Rate}
Studies have shown that drinking fluoridated water can keep teeth strong and reduce cavities by about 25\% in children and adults.\footnote{\url{https://www.cdc.gov/fluoridation/index.html}} To better understand how community water fluoridation rate influence the dental health topics that people usually tweet about, we used the latest state-level fluoridation statistics from the Centers of Disease Control and Prevention to approximate the fluoridation rate of water that users use and drink.\footnote{\url{https://www.cdc.gov/fluoridation/statistics/2018stats.htm/}}

\subsubsection{Pandemic Severity}
To measure the pandemic severity, we calculated the county-level 7-day average relative change in the number of daily COVID-19 confirmed cases. Each user was associated with a tweet in our dataset, and there was a timestamp for each tweet. We chose the date the user posted the tweet to calculate this variable. The data was collected from the data repository maintained by the Center for Systems Science and Engineering (CSSE) at John Hopkins University~\cite{dong2020interactive}.

\begin{figure}[t]
    \centering
    \includegraphics[width = \linewidth]{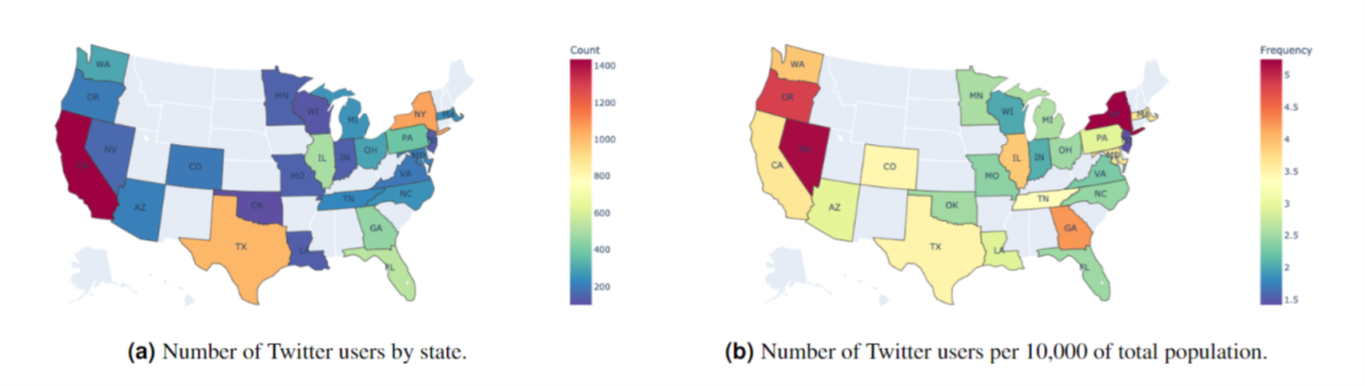}
    \caption{State-level user distributions.}
    \label{fig:state_level}
\end{figure}

\begin{figure}[t]
    \centering
    \includegraphics[width = \linewidth]{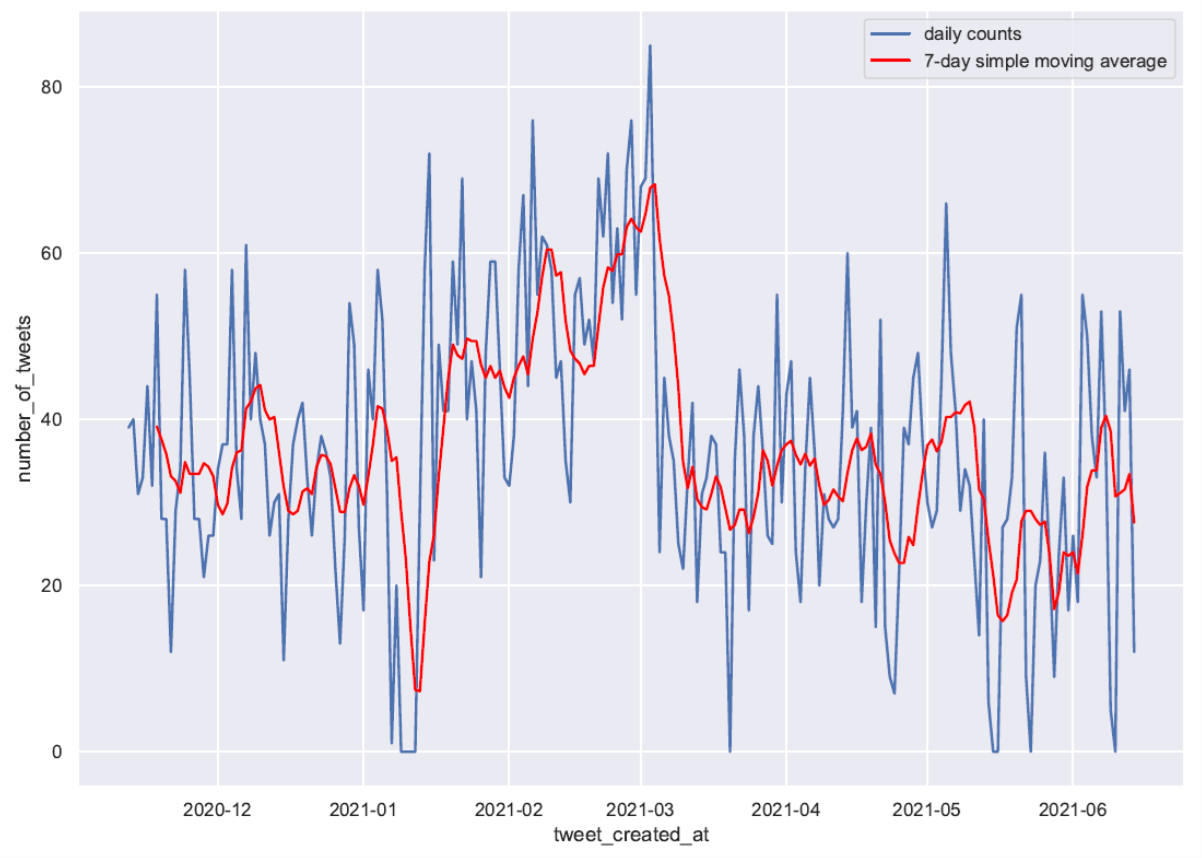}
    \caption{Trend of tweet activity (unique tweets).}
    \label{fig:trend}
\end{figure}

After inferring or extracting gender, age, population density, income, poverty rate, health insurance coverage rate, community fluoridation rate, and relative change of the number of daily confirmed COVID-19 cases and retaining states having at least 100 unique users, our final dataset consists of 10,883 tweets posted by 9,104 Twitter users with all inferred features included. We excluded states with fewer than 100 unique users to ensure our data quality, i.e., by removing noises from states with small sample sizes. Figure~\ref{fig:state_level}a shows the geographic distribution of Twitter users in our study. California, Texas, and New York are the top three states with regard to the number of users who tweet about oral health. However, this could be because that these states are most active in general.\footnote{\url{https://www.allbusiness.com/twitter-ranking-which-states-twitter-the-most-12329567-1.html}} Figure~\ref{fig:state_level}b shows that New York, Nevada, and Oregon tend to have higher relative frequency of users who tweet about oral health. It is noteworthy that Oregon has the highest dental care utilization among adults with private dental benefits.\footnote{\url{http://www.ada.org/~/media/ADA/Science\%20and\%20Research/HPI/OralHealthCare-StateFacts/Oral-Health-Care-System-Full-Report.pdf}} Figure~\ref{fig:trend} illustrates the trend of daily frequency of unique tweets. Apart from the big “down” and “up” between early January 2021 and mid-March 2021 which roughly correspond to the trend of daily COVID-19 confirmed cases in the US~\cite{dong2020interactive}, the daily tweet activity stays relatively stable and varies mostly between 20 and 60 tweets per day.

\subsection{Topic Modeling}
We used LDA~\cite{blei2003latent} to extract topics from tweets. We removed Twitter handles, hashtags, links, punctuation, numbers, special characters, and stop words to clean the texts of our tweets. We used the spaCy package to only keep the words whose postag is either “NOUN”, “ADJ”, “VERB”, or “ADV”. We applied grid search to find the optimal hyperparameters setting for num\_topics, $\alpha$, and $\beta$. The optimal setting is as follows: num\_topics=5, $\alpha$=0.01, and $\beta$=0.41, with a coherence score of 0.53.

\subsection{Logistic Regression}
We constructed logistic regression models for five topics uncovered by topic modeling. The target for logistic regression model $LR_i$ is binary with label 1 indicating that a user has posted a tweet $\in Topic_i$ and label 0 indicating that a user has not posted a tweet $\in Topic_i$. Coefficients of features in $LR_i$ are used to interpret the strength and direction of the associations between Twitter users' individual features and $Topic_i$. In our dataset, on average, one user posted 1.19 tweets during the entire study period. We could easily assign the topic of each user based on the single tweet he/she posted. However, it is still possible that a user posted multiple tweets of different topics. We assigned the topic that a user posted most to be his/her topic. In the end, the study group is composed of 9,104 unique Twitter users. The choices of the topics might not be independent. To verify this, we randomly sampled 100 users and read the corresponding tweets. We found that only 5 tweets were the reactions to other users’ posts. 

\section{Results}
To address RQ1, we attempted to capture what topics or oral diseases are discussed when Twitter users post about oral health. Table~\ref{tab:lda_topics_summary} shows the five topics extracted by the LDA topic modeling. We assigned the title of each topic based on its top ten keywords. One word might appear in multiple topics, but the rank of the word in that topic indicates the importance of it to the topic. Guided by the combinations of the keywords, we inferred the focus of each topic. For instance, “cavity” and “dentistry” are in Topic 2. “gum”, “decay”, “bleed”, “food” are in Topic 3. By reading the keywords, we knew the focuses of these two topics are different. For simplicity, we summarized the topics with short labels. However, to interpret the meaning of each topic, we need to refer to the keywords instead of the labels.

\begin{table*}[t]
\caption{Top 10 keywords of each topic.}
    \label{tab:lda_topics_summary}
    \centering
    \begin{tabular}{l l l}
    \hline
        Topic & Keywords \\
         \hline
        (1) Dental pain & pain, dental, work, good, oral, surgery, care, start, lot, people \\
        (2) Dental service/cavity & dentist, cavity, time, make, year, dentistry, ago, fill, month, call\\
        (3) Tooth decay/gum bleeding & tooth, gum, decay, eat, bleed, food, brush, hard, stop, floss\\
        (4) Wisdom tooth pain/jaw hurt & pain, wisdom, bad, tooth, hurt, day, jaw, mouth, week, back\\
        (5) Chipped tooth/tooth break & tooth, chip, break, feel, today, pull, give, front, fix, toothache\\
    \hline
    \end{tabular}
\end{table*}

\begin{figure}[t]
    \centering
    \includegraphics[width = \linewidth]{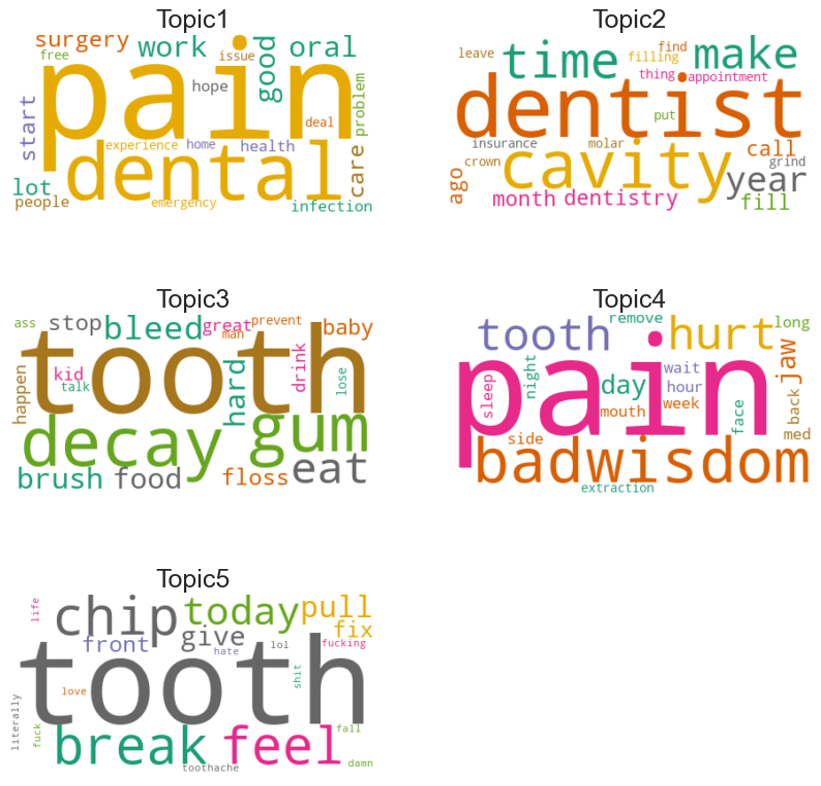}
    \caption{Word cloud of the five topics extracted by the LDA model.}
    \label{fig:wordcloud}
\end{figure}

We visualized the relative frequency of keywords in each topic in Figure~\ref{fig:wordcloud}. The differences of the top keywords between topics suggest that LDA has captured the major component of each topic, and there is a good separation. LDA calculated the weights of five topics of each tweet. The topic that has the highest weight is considered as the dominant topic of the tweet. We grouped the tweets into five classes based on their dominant topic.

As indicated by the previous studies, people increased consumption of products that are detrimental to oral health during work from home.\footnote{\url{https://www.myfooddata.com/articles/high-sugar-foods.php}} To investigate such effects, we constructed a list of keywords of snacks and alcohol (“drunk”, “liquor”, “beer”, “champagne”, “wine”, “gin”, “vodka”, “rum”, “whisky”, “brandy”, “tequila”, “bourbon”, “sweet food”, “sugar”, “candy”, “spaghetti sauce”, “sports drinks”, “chocolate milk”, “granola”, “honey”, “glucose”, “corn sugar”, “milkshakes”, “juice”, “cream soda”, “cake”, “cereal”, “chocolate”, “honey”, “milk”, “yogurt”, “ice cream”, “cookie”, “dried sweetened mango”, “candied tamarind”) and performed a keyword search in all the tweets to examine whether or not the tweets mention sweet snacks/drinks or alcohol. Overall, 1.7\% tweets mention sweets or alcohol. In particular, the tweets mentioning sweets or alcohol account for 1.0\%, 0.4\%, 4.2\%, 0.7\% and 1.8\% for the topic of Dental pain, Dental service/cavity, Tooth decay/gum bleeding, Wisdom tooth pain/jaw hurt and Chipped tooth/tooth break, respectively. A higher proportion of tweets mentioning sweets or alcohol was observed in Tooth decay/gum bleeding and Chipped tooth/tooth break. For each user, we assigned the topic that the user tweets most frequently as the dominant topic. The proportions of the topics of users are as follows (Table~\ref{tab:level1_logit_sum}):

\begin{table*}[htbp]
\centering
\begin{threeparttable}
\caption{Topic distribution by user characteristics.}
    \label{tab:level1_logit_sum}
    \centering
    
    \begin{tabular}{l l l l l l}
    \hline
    Variable     & Dental pain & Dental service & Tooth decay & Wisdom tooth pain & Chipped tooth \\
     &  & /cavity  & /gum bleeding  & /jaw hurt & /tooth break \\
      & (1) & (2) & (3) & (4) & (5) \\
         \hline
        \textbf{Total}  & \makecell[l]{16.23\%} & \makecell[l]{23.86\%} & 
        \makecell[l]{14.24\%} & 
        \makecell[l]{26.70\%} & 
        \makecell[l]{18.97\%}\\
    
        Urban  & \makecell[l]{16.01\%} & \makecell[l]{24.68\%} & 
        \makecell[l]{13.77\%} & 
        \makecell[l]{26.66\%} & 
        \makecell[l]{18.88\%}\\
        
        Suburban & \makecell[l]{17.00\%} & \makecell[l]{24.23\%} & 
        \makecell[l]{16.37\%} & 
        \makecell[l]{24.50\%} & 
        \makecell[l]{17.90\%}\\

        Rural & \makecell[l]{16.67\%} & \makecell[l]{19.85\%} & 
        \makecell[l]{14.73\%} & 
        \makecell[l]{28.56\%} & 
        \makecell[l]{20.19\%}\\
      
        Age $\leq$ 18 & \makecell[l]{10.74\%} & \makecell[l]{24.16\%} & 
        \makecell[l]{14.09\%} & 
        \makecell[l]{34.90\%} & 
        \makecell[l]{16.11\%}\\
    
        Age 19-29 & \makecell[l]{12.83\%} & \makecell[l]{26.99\%} & 
        \makecell[l]{11.38\%} & 
        \makecell[l]{28.12\%} & 
        \makecell[l]{20.67\%}\\
    
        Age 30-49 & \makecell[l]{18.12\%} & \makecell[l]{21.30\%} & 
        \makecell[l]{15.09\%} & 
        \makecell[l]{27.21\%} & 
        \makecell[l]{18.27\%}\\
      
        Age 50-64 & \makecell[l]{23.38\%} & \makecell[l]{19.27\%} & 
        \makecell[l]{22.68\%} & 
        \makecell[l]{19.74\%} & 
        \makecell[l]{14.92\%}\\
       
        Age $\geq$ 65 & \makecell[l]{28.42\%} & \makecell[l]{19.06\%} & 
        \makecell[l]{23.38\%} & 
        \makecell[l]{14.75\%} & 
        \makecell[l]{14.39\%}\\
      
        Male & \makecell[l]{16.34\%} & \makecell[l]{20.13\%} & 
        \makecell[l]{16.73\%} & 
        \makecell[l]{27.42\%} & 
        \makecell[l]{19.38\%}\\

        Female & \makecell[l]{16.16\%} & \makecell[l]{26.63\%} & 
        \makecell[l]{12.39\%} & 
        \makecell[l]{26.17\%} & 
        \makecell[l]{18.66\%}\\

    \hline
    \end{tabular}
    \begin{tablenotes}
    \item
    \end{tablenotes}
    \end{threeparttable}
\end{table*}

Topic 1 (Dental pain) accounts for 16.23\% of all Twitter users. In these tweets, people often mention pains caused by dental diseases or infections. An example tweet is:

{\tt I've been dealing with severe dental pain for the past like 10 years. I can handle it but it suuuucks.  Compounded that I just had another molar yanked out on Monday morning.}

Topic 2 (Dental service/cavity) represents 23.86\% of all Twitter users in our study, where people mainly share their experience with dentists to fix their dental problems and/or talk about specific oral disease like cavity. An example tweet is:

{\tt Shout out to Smile Studio in Zachary. I hate going to the dentist, and I am super-sensitive to pain. 44 years without a cavity and I had to have an extraction today. They made it almost painless.}

Topic 3 (Tooth decay/gum bleeding) constitutes 14.24\% of all users, including the keywords “tooth”, “gum”, “decay”, “eat”, and “bleed”. In this topic, Twitter users mainly talk about two of the most common oral diseases - tooth decay and gum bleeding. An example tweet is:

{\tt My fake tooth chipped off on Xmas one month after losing dental so I’m ugly now 29. 62\% of people who brush their teeth rinse their mouth out with water, which actually makes tooth decay more likely.}

Topic 4 (Wisdom tooth pain/jaw hurt) is the most tweeted topic in our study, which accounts for 26.70\% of all users. People mostly post about pains from wisdom tooth or jaw hurt. It contains keywords “chip”, “front”, “dog”, “miss”, and “walk”. An example tweet is:

{\tt Yooo I don’t wish the wisdom tooth pain not even on my worse enemy. Shit is wild Painful tooth Can't get straight to your dentist $<hashtag>$ $<hashtag>$.}

Topic 5 (Chipped tooth/tooth break) represents 18.97\% of all users, where people mostly talk about chipped tooth/teeth or break their own tooth/teeth.  The keywords include “tooth”, “chip”, “break”, “feel”, and “today”. An example tweet is:

{\tt I’m gonna get my chipped front tooth fixed tomorrow. It was broken when my dad smacked me as a kid. His wedding ring. He regretted it. I probably deserved it. I’ve been getting it fixed for 30 years. God, I miss him.}

\subsection{Logistic Regression Results}
\begin{figure}[t]
    \centering
    \includegraphics[width = \linewidth]{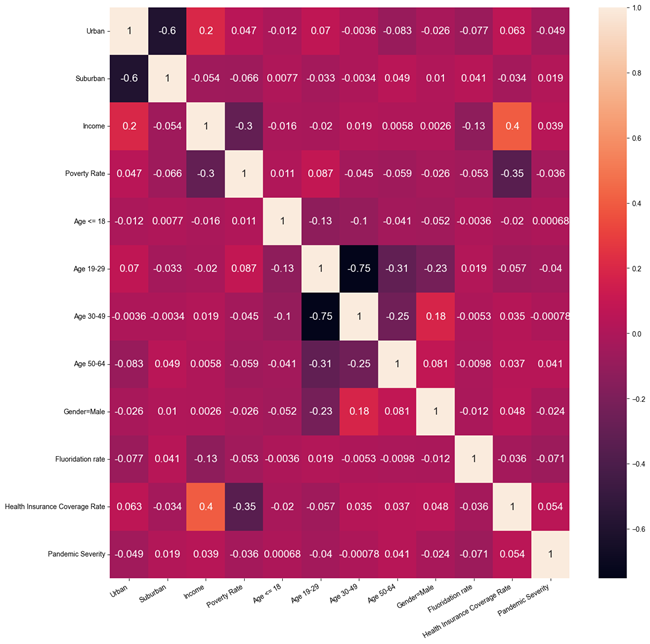}
    \caption{Correlation heat map of the variables for logistic regression (Best viewed by zoom-in on screen).}
    \label{fig:correlation}
\end{figure}

To answer RQ2 and RQ3, we conducted multiple logistic regression analyses to examine how each variable including age, gender, population density, income and poverty rate influence whether or not a user will tweet about a specific topic. Table~\ref{tab:logit_result} lists the logistic regression results for different topics against variables of interest. Each column represents a logistic regression model. Figure~\ref{fig:correlation} illustrates the correlations between the variables for logistic regression analyses.

\begin{table*}[htbp]
\centering
\begin{threeparttable}
\caption{Logistic regression outputs for all topics against variables of interest.}
    \label{tab:logit_result}
    \centering
    
    \begin{tabular}{l l l l l l}
    \hline
    Variable     & Dental pain & Dental service & Tooth decay & Wisdom tooth pain & Chipped tooth \\
     &  & /cavity  & /gum bleeding  & /jaw hurt & /tooth break \\
      & (1) & (2) & (3) & (4) & (5) \\
         \hline
        Urban (0=No, 1=Yes) & 
        \makecell[l]{-0.0045 \\ (0.081)} & \makecell[l]{$0.1550^{*}$ \\ (0.074)} & \makecell[l]{0.0334 \\ (0.086)} & 
        \makecell[l]{-0.1278 \\ (0.067)} & 
        \makecell[l]{-0.0588\\(0.075)}\\
        
        Suburban (0=No, 1=Yes)& 
        \makecell[l]{-0.0001 \\ (0.108)} &  \makecell[l]{$0.2044^{*}$ \\ (0.097)} & \makecell[l]{0.1419 \\ (0.111)} & 
        \makecell[l]{-$0.2290^{*}$ \\ (0.092)} & 
        \makecell[l]{-0.1435\\(0.102)}\\
        
        Income & 
        \makecell[l]{0.0316 \\ (0.027)} &  \makecell[l]{$0.0979^{***}$ \\ (0.022)} & \makecell[l]{-0.0078 \\ (0.030)} &
        \makecell[l]{-0.0326 \\ (0.024)} & 
        \makecell[l]{-$0.0664^{*}$ \\ (0.028)} \\
        
        Poverty Rate & 
        \makecell[l]{-$0.0222^{**}$ \\ (0.007)} &  \makecell[l]{$0.0202^{***}$ \\ (0.005)} & 
        \makecell[l]{-$0.0138^{*}$ \\ (0.007)} & \makecell[l]{-$0.0139^{**}$ \\ (0.005)} & 
        \makecell[l]{-0.0048 \\ (0.006)}\\
        
        Age $\leq$ 18 (0=No, 1=Yes)& \makecell[l]{-$1.2473^{***}$ \\ (0.296)} & \makecell[l]{0.0010 \\ (0.240)} & \makecell[l]{-0.4958 \\ (0.275)} & \makecell[l]{$1.0641^{***}$ \\ (0.235)} & 
        \makecell[l]{0.0609 \\ (0.275)}\\
        
        Age 19-29 (0=No, 1=Yes)& \makecell[l]{-$1.0320^{***}$ \\ (0.139)} &  \makecell[l]{0.1600 \\ (0.149)} & \makecell[l]{-$0.7450^{***}$ \\ (0.149)} & \makecell[l]{$0.7446^{***}$ \\ (0.163)} &
        \makecell[l]{$0.3788^{*}$ \\ (0.165)}\\
        
        Age 30-49 (0=No, 1=Yes)& \makecell[l]{-$0.6085^{***}$ \\ (0.138)} & \makecell[l]{-0.0605 \\ (0.150)} & \makecell[l]{-$0.4925^{**}$ \\ (0.148)} & \makecell[l]{$0.6735^{***}$ \\ (0.164)} & 
        \makecell[l]{0.2028 \\ (0.166)}\\
        
        Age 50-64 (0=No, 1=Yes)& 
        \makecell[l]{-0.3000 \\ (0.154)} &  \makecell[l]{-0.1474 \\ (0.168)} & 
        \makecell[l]{-0.0189\\ (0.162)} & 
        \makecell[l]{0.2632 \\ (0.181)} & 
        \makecell[l]{-0.0519 \\ (0.187)}\\
        
        Male (0=No, 1=Yes)& 
        \makecell[l]{-$0.1291^{*}$ \\ (0.060)} &  \makecell[l]{-$0.3082^{***}$ \\ (0.053)} & \makecell[l]{$0.2700^{***}$ \\ (0.062)} &
        \makecell[l]{$0.1077^{*}$ \\ (0.050)} &
        \makecell[l]{0.1083 \\ (0.056)}\\
        
        Fluoridation Rate & 
        \makecell[l]{-0.0010 \\ (0.002)} & \makecell[l]{-0.0005 \\ (0.001)} & 
        \makecell[l]{-0.0008 \\ (0.002)} &
        \makecell[l]{-0.0014 \\ (0.001)} & 
        \makecell[l]{0.0005 \\ (0.002)}\\
        
        Health Insurance Coverage Rate& 
        \makecell[l]{-$0.0071^{**}$ \\ (0.003)} &
        \makecell[l]{-$0.0184^{***}$ \\ (0.002)} & 
        \makecell[l]{-$0.0147^{***}$ \\ (0.003)} & \makecell[l]{-$0.0119^{***}$ \\ (0.002)} & \makecell[l]{-$0.0169^{***}$ \\ (0.003)}\\
        
        Pandemic Severity& 
        \makecell[l]{0.0465 \\ (0.049)} &
        \makecell[l]{-$0.1730^{***}$ \\ (0.048)} & 
        \makecell[l]{$0.2824^{***}$ \\ (0.048)} & \makecell[l]{-$0.2760^{***}$ \\ (0.047)} & \makecell[l]{$0.1718^{***}$ \\ (0.045)}\\
        
        N &  \multicolumn{5}{c}{9,104}\\
    \hline
    \end{tabular}
    \begin{tablenotes}
    \item Note. * $p<0.05$. ** $p<0.01$. *** $p<0.001$. Table entries are coefficients (standard errors). Income is scaled down by 10,000 while poverty rate, fluoridation rate, pandemic severity, and  health insurance coverage rate are scaled up by 100.
    \end{tablenotes}
    \end{threeparttable}
\end{table*}

\textbf{Older adults tend to tweet more about dental pain but less about wisdom tooth pain/jaw hurt.} In logistic regression analysis, we divided age into five groups: $\leq$ 18, 19-29, 30-49, 50-64, and $\geq$ 65. As for the topic of tooth pain/dental services, logistic regression coefficients for the first three age groups are all negative, which means adults in their 50s or above are more likely to talk about dental pain in general ($p < .001$). As for the topic of wisdom tooth pain/jaw hurt, logistic regression coefficients for the first three groups are positive, which means adults in their 50s or above are less likely to talk about wisdom tooth pain/jaw hurt ($p < .001$).

\textbf{Women tend to talk more about dental pain and dental service/cavity while men tweet more tooth decay/gum bleeding and wisdom tooth pain/jaw hurt.} We found men are less likely to talk about dental pain ($B = -0.1291, SE = 0.060, p < .05, OR = 0.8789, 95\% CI = [0.7819,0.9881]$) and dental service/cavity ($B = -0.3082, SE = 0.053, p < .001, OR = 0.7402, 95\% CI = [0.6630,0.8146]$). Men are more likely to talk about tooth decay/gum bleeding ($B = 0.2700, SE = 0.062, p < .001, OR = 1.3100, 95\% CI = [1.1595,1.4800]$) and wisdom tooth pain/jaw hurt ($B = 0.1077, SE = 0.050, p < .05, OR = 1.1137, 95\% CI = [1.0101,1.2275]$).

\textbf{People from rural areas are less likely to discuss dental service/cavity and people from suburban areas are less likely to talk about wisdom tooth pain/jaw hurt.} By conducting logistic regression, we found that the Twitter users who live in rural areas tweet less about dental service/cavity ($p < .05$) since the coefficients for urban and suburban are both positive in logistic regression analysis of the topic dental service/cavity. We found that suburban people tweet less about wisdom tooth pain/jaw hurt ($B = -0.2290, SE = 0.092, p < .05, OR = 0.7953, 95\% CI = [0.6643,0.9522]$).

\textbf{Higher-income people tend to talk more about dental service/cavity but less about chipped tooth/tooth break.} If the income increases by 10,000, the odds of tweeting about dental service/cavity is 1.1029 times ($B = 0.0979, SE = 0.022, p < .001, OR = 1.1029, 95\% CI = [1.0555,1.1526]$) and the odds of tweeting about chipped tooth/tooth break is 0.9358 times ($B = -0.0664, SE = 0.028, p < .05, OR = 0.9358, 95\% CI = [0.8869,0.9881]$).

\textbf{People from counties with higher poverty rate talk less about dental pain, tooth decay/gum bleeding, and wisdom tooth pain/jaw hurt and more about dental service/cavity.} If the poverty rate increase by 1\%, the odds of talking about dental pain is 0.9780 times ($B = -0.0222, SE = 0.007, p < .01, OR = 0.9780, 95\% CI = [0.9656,0.9910]$), the odds of talking about tooth decay/gum bleeding is 0.9863 times ($B = -0.0138, SE = 0.007, p < .05, OR = 0.9863, 95\% CI = [0.9734,0.9995]$), the odds of talking about wisdom tooth pain/jaw hurt is 0.9862 times ($B = -0.0139, SE = 0.005, p < .01, OR = 0.9862, 95\% CI = [0.9763,0.9960]$), and the odds of talking about dental service/cavity is 1.0204 times ($B = 0.0202, SE = 0.005, p < .001, OR = 1.0204, 95\% CI = [1.0101,1.0315]$).

\textbf{People from counties with higher health insurance coverage rate tend to tweet less about all oral health-related topics.} If the health insurance coverage rate increases by 1\%, the odds of tweeting about dental pain is 0.9929 times ($B = -0.0071, SE = 0.003, p < .01, OR = 0.9929, 95\% CI = [0.9881,0.9980]$), dental service/cavity is 0.9817 times ($B = -0.0184, SE = 0.002, p < .001, OR = 0.9817, 95\% CI = [0.9773,0.9861]$), tooth decay/gum bleeding is 0.9854 times ($B = -0.0147, SE = 0.003, p < .001, OR = 0.9854, 95\% CI = [0.9802,0.9910]$), wisdom tooth pain/jaw hurt is 0.9882 times ($B = -0.0119, SE = 0.002, p < .001, OR = 0.9882, 95\% CI = [0.9831,0.9930]$), and chipped tooth/tooth break is 0.9832 times ($B = -0.0169, SE = 0.003, p < .001, OR = 0.9832, 95\% CI = [0.9782,0.9881]$).

\textbf{People from counties at a higher risk of COVID-19 talk less about dental service/cavity, wisdom tooth pain/jaw hurt and more about tooth decay/gum bleeding and chipped tooth/tooth break.} If the 7-day average relative change of the number of daily COVID-19 confirmed cases grows by 1\%, the odds of tweeting about dental service/cavity is 0.8411 times ($B = -0.1730, SE = 0.048, p < .001, OR = 0.8411, 95\% CI = [0.7657,0.9240]$), wisdom tooth pain/jaw hurt is 0.7588 times ($B = -0.2760, SE = 0.047, p < .001, OR = 0.7588, 95\% CI = [0.6921,0.8319]$), tooth decay/gum bleeding is 1.3263 times ($B = 0.2824, SE = 0.048, p < .001, OR = 1.3263, 95\% CI = [1.2080,1.4564]$), and chipped tooth/tooth break is 1.1874 times ($B = 0.1718, SE = 0.045, p < .001, OR = 1.1874, 95\% CI = [1.0876,1.2969]$).

\section{Discussion}
Among the Twitter users in our study, 26.70\% talk about wisdom tooth pain/jaw hurt, 23.86\% tweet about dental service/cavity, 18.97\% discuss chipped tooth/tooth break, 16.23\% talk about dental pain, and 14.24\% talk about tooth decay/gum bleeding. Overall, women are more likely to discuss oral health amid the COVID-19 pandemic. On one hand, this might be because men are more likely to ignore their dental health and visit dentists less frequently for disease prevention~\cite{lipsky2021men}. On the other hand, studies~\cite{stouthard1993assessment,doerr1998factors} showed that women are more dentally anxious, which might lead to physiological, cognitive, behavioral, health, and social issues~\cite{cohen2000impact}. COVID-19 has also changed people's work patterns as many companies encourage or require employees to work from home to prevent the spread of virus,\footnote{\url{https://www.gartner.com/en/newsroom/press-releases/2020-03-19-gartner-hr-survey-reveals-88--of-organizations-have-e}} which was found to influence people's oral health by increasing the consumption of products that are detrimental to oral health such as snacks and alcohol and increasing the consumption of oral health products.\footnote{\url{https://bridgedental.com/2020/04/26/the-impact-of-working-from-home-during-the-coronavirus-lockdown-on-dental-health/}} Another potential reason that women tend to talk about oral health amid the COVID-19 pandemic is that they are more likely to reduce work hours and spend more time on oral health since they can stay home longer~\cite{collins2021covid,xiong2021social}. With respect to age, younger adults (19-29) tend to tweet more often about oral health problems. This echoes the finding that younger adults experience a higher prevalence of dental fear and anxiety (DFA), high DFA, and severe DFA~\cite{silveira2021estimated}.  

We observed that the topics of interest vary across user characteristics including age, gender, population density, income, poverty rate, and health insurance coverage rate. Older adults, who are identified as the highest risk group for fatal COVID-19 clinical outcomes~\cite{marchini2020covid,sharma2021estimating} are more likely to talk about dental pain ($p < .001$). Due to the pandemic, older adults are facing lack of access to the oral health care~\cite{wall2012recent}. We also found the older adults are less likely to tweet about wisdom tooth pain/jaw hurt which is within our expectation as the age at which the wisdom teeth erupt is generally early twenties~\cite{sawyer2018age}. It is noteworthy that we intended to focus on discovering the age pattern instead of any specific age group. According to the logistic regression results shown in Table~\ref{tab:logit_result}, the coefficients of the age groups that passed the statistical tests have shown a \textit{consistent} pattern with the ages. That is, the changes in the coefficients are consistent with the changes in the ages.

Women tend to focus more on dental pain ($p < .05$) and dental service/cavity ($p < .001$) while men are more interested in discussing tooth decay/gum bleeding ($p < .001$) and wisdom tooth pain/jaw hurt ($p < .05$). Studies~\cite{furuta2011sex,heft2007gender} showed that women are almost twice as likely to have received a regular dental check-up and are more likely to report general fear of dental pain compared to men.

People from rural areas are less likely to discuss dental service/cavity ($p < .05$) which is possibly due to the lack of access to dental care. People living in rural America have about 8\% (children) to 10\% (adults aged 18–64) less access to dental services compared with their urban counterparts in 2017.\footnote{\url{https://www.cdc.gov/nchs/data/hus/2018/037.pdf}} Suburban people talk less about wisdom tooth pain/jaw hurt ($p < .05$). 

Higher-income people talk more about dental service/cavity ($p < .001$) but less about chipped tooth/tooth break ($p < .05$). High cost is the most significant reason for people not visiting dentists in the US.\footnote{\url{https://www.ada.org/~/media/ADA/Science\%20and\%20Research/HPI/OralHealthWell-Being-StateFacts/US-Oral-Health-Well-Being.pdf}} Higher-income people may care less about the high cost and have more frequent dental services which suggests the disparities in oral health. People from counties with a higher poverty rate talk less about dental pain ($p < .01$), tooth decay/gum bleeding ($p < .05$), and wisdom tooth pain/jaw hurt ($p < .01$). 

Although \citet{kim2017associations} suggested that Community Water Fluoridation is associated with lowering the risks of having certain oral diseases like dental caries, our study has shown that state-level fluoridation rate is not associated with the prediction of any oral health-related topics. This may be due to the limitation of not having publicly available more granular fluoridation rate data. 

Health insurance coverage rate is the most important predictor for the logistic regression for topic prediction. People from counties with a higher insurance coverage rate tend to tweet less about all topics of oral health ($p < .01$) which is consistent with the findings of \citet{zivkovic2020providing} that health or more specifically dental insurance plays an important role in improving people’s oral health conditions. Combining with the findings with respect to the income and poverty rate, we think these variables are related with oral health because they impact the ability to pay for dental service, preventive medical and dental care. 

With respect to the pandemic severity, people from counties at a higher risk of COVID-19 talk less about dental service/cavity ($p < .001$) and wisdom tooth pain/jaw hurt ($p < .001$) but more about tooth decay/gum bleeding ($p < .001$) and chipped tooth/tooth break ($p < .001$). On one hand, it is likely that people delay or avoid dental visits because of closure and reduced hours of dental care and the fear of being infected with the virus during their dental appointments~\cite{zachary2020oral,wong2020all}. On the other hand, COVID-19 has a negative effect on oral health possibly resulting from xerostomia, loss of taste or smell sensation, and mental health breakdown~\cite{farook2020covid}.

There might be potential biases in the study samples. We chose to conduct our study at a large scale with almost ten thousand participants to address the bias issue. More importantly, as we discussed above, most of our findings are \textit{consistent} with the previous studies and polls. It further validates the robustness of our study. There is another merit of our study in addressing the sample biases. Collecting social media data, which is observing the social media users in a passive way, could potentially alleviate the social desirability biases~\cite{krumpal2013determinants} introduced in the survey data of a traditional design~\cite{mokdad2010measuring}.

The findings of our study have implications for the community of general dentists. The public social media can be an important source of information for the community to have a more comprehensive understanding of the dental need, especially during a pandemic when the accessibility to the dental services is limited and the communication between the patients and the dentists is less frequent. Policies can be designed to provide more dental care to the people in need based on the finding that the topics of interest vary across user characteristics.

\section{Limitations}
There are some limitations of our study. Some states with a smaller sample size are not included in the study population. Temporal changes have not been investigated. Future work can be directed to analyzing oral health-related discussions across multiple social media platforms and combining the insights of the survey data to achieve broader and more comprehensive perspectives.

\section{Conclusion}
This is the first large-scale social media-based study to understand the public discussions on oral health during the COVID-19 pandemic in the US. We have analyzed 10,883 tweets from 9,104 Twitter users across 26 states (with sufficient samples) during the period of November 12, 2020 to June 14, 2021. The topics of interest vary across user characteristics. Tweets inform social disparities in oral health during the pandemic. We hope our work can promote research on public health issues through the lens of social media, provide insights for oral health practitioners and policy makers, enhance the public awareness of the importance of oral health, and ultimately improve oral health in America amid the COVID-19 pandemic.


\bibliography{sample-base}

\end{document}